\documentstyle[11pt]{article}

\def\d{{\partial}}
\newcommand{\be}{\begin{equation}}
\newcommand{\ee}{\end{equation}}
\newcommand{\ba}{\begin{eqnarray}}
\newcommand{\ea}{\end{eqnarray}}
\newcommand{\no}{\nonumber}
\newcommand{\al}{\alpha}
\newcommand{\bt}{\beta}

\newcommand{\la}{\langle}
\newcommand{\ra}{\rangle}
\newcommand{\Op}{{\cal O}}
\newcommand{\e}{\epsilon}

\begin{document}

\thispagestyle{empty}

\begin{flushright}
UT-789, YITP-97-47\\
September, 1997 \\
\end{flushright}

\bigskip

\begin{center}
{\Large \bf Quantum Cohomology
and Free Field Representation}
\end{center}

\bigskip

\bigskip

\begin{center}

Tohru Eguchi 

\bigskip

{\it Department of Physics, University of Tokyo, 

\medskip

Tokyo 113, Japan}

\bigskip

\bigskip

Masao Jinzenji

\bigskip

{\it Department of Mathematical Science, University of Tokyo

\medskip

Tokyo 153, Japan}

\bigskip

\bigskip

and

\bigskip

\bigskip

Chuan-Sheng Xiong

\medskip

{\it Yukawa Institute for Theoretical Physics, Kyoto University 

\bigskip

Kyoto 606, Japan}
\end{center}

\bigskip

\bigskip

\begin{abstract}
In our previous article we have proposed that the Virasoro algebra controls
the quantum cohomology of Fano varieties at all genera. In this paper
we construct a free field description of Virasoro operators and quantum
cohomology. We shall show that to each even (odd) homology class of a 
K\"{a}hler manifold we have a free bosonic (fermionic) field and Virasoro 
operators are given by a simple bilinear form of these fields. 
We shall show that the Virasoro condition correctly reproduces the 
Gromov-Witten invariants also in the case of manifolds
with non-vanishing non-analytic classes ($h^{p,q}\not=0,p\not=q$)
and suggest that the Virasoro condition holds universally for all compact 
smooth K\"{a}hler manifolds. 
\end{abstract}

\newpage
\pagenumbering{arabic}

\noindent{\large \bf 1. Introduction} \\

\bigskip

In our previous communication we have proposed a new approach to
the quantum cohomology theory: control of the structure of quantum 
cohomology at all genera by the Virasoro algebra \cite{EHXa}. 
As we know, the Virasoro algebra appeared before in the study of
2-dimensional gravity \cite{DVV} and the intersection theory on the
moduli space of Riemann surfaces \cite{Wa, Konts}. 
Since the quantum cohomology theory 
describes the interplay between the geometry of
the target manifold and moduli space of Riemann surfaces, it is not
surprising that the Virasoro algebra appears here also. In \cite{EHXa}
we have shown that the Virasoro conditions when applied to projective spaces 
and Grassmanianns reproduce exactly the known Gromov-Witten
invariants at genus $g=0$ and $1$ \cite{KM,Itz,JS,CH,Getz,Pand}.

In this paper we would like to 
discuss a free-field description of the Virasoro algebra and quantum
cohomology theory. As we shall see, given an arbitrary
K\"{a}hler manifold $M$ there exists a 
2-dimensional free bose (fermi) field 
for each even (odd)-dimensional homology class of $M$. 
These fields have non-zero pairings (OPE's) when their corresponding
homology cycles have non-zero intersections in $M$. 
Virasoro operators are given by a simple bilinear form of these fields.
The central charge of the Virasoro algebra equals the (the number 
of bosons) - (the number of fermions (ghosts)) = the Euler characteristic of 
$M$. 

It seems to us rather
striking that such a correspondence exists between
2-dimensional conformal fields and homology classes of
K\"{a}hler manifolds. Our observation
may be considered as a natural generalization
of the correspondence between bosonic (fermionic) vacua and
even (odd)-dimensional homology cycles
in supersymmetric quantum mechanics. Vacua are replaced by free fields and 
the Witten index is replaced by the Virasoro central charge.

In this paper we also would like to discuss the case of 
K\"{a}hler manifolds $M$ where the "off-diagonal" cohomology classes 
$H^{p,q},\, p\not=q$ are present. As proposed by S.Katz \cite{Kz}, in such a
case the original expressions of the Virasoro operators in \cite{EHXa}
should be modified and the "holomorphic dimension" $p$ (or "anti-holomorphic 
dimension" $q$) should be used for the dimensions of the fields 
in the Virasoro operators. In fact this proposal may be verified if we repeat
the orginal derivation of Virasoro condition by carefully
treating the ghost number conservation relation which holds for its
holomorphic and anit-holomorphic parts separately.
We have examined Katz's proposal 
and have verified that it reproduces the correct
instanton numbers at genus $g=0$ and $1$ in the case of cubic 
hypersurface in $CP^4$. \\

\bigskip

\noindent{\large \bf 
2. Complex Projective Spaces and Free Field Representation} \\

\bigskip

Let us first consider, for simplicity, the case of complex projective spaces
$CP^N \, (N=\mbox{even})$. Corresponding to each cohomology class
$H^{q_{\al},q_{\al}} \, (q_{\al}=0,1,\cdots,N)$ of $CP^N$, 
one introduces an infinite 
set of parameters $\{t_{\al,n},\,n\ge 0\}$ which describe the primary ($n=0$)
and descendant ($n\ge 1$) couplings.
The Virasoro operators are given by
\ba
&&\hskip-8mm L_n=\sum_{m=0}^\infty\sum_{\alpha=0}^N 
\sum_{j=0}^{N-\alpha}
(N+1)^jC_\alpha^{(j)}(m,n) t_m^\alpha\d_{\alpha+j,m+n-j} \hskip20mm n\ge 1
\label{ln}\\
&&\hskip-5mm+\frac{\lambda^2}{2}\sum_{\alpha=0}^N
\sum_{j=0}^{N-\alpha}\sum_{m=0}^{n-j-1}
(N+1)^jD_\alpha^{(j)}(m,n) \d^\alpha_m\d_{\alpha+j,n-m-j-1}
+\frac{1}{2\lambda^2}\sum_{\alpha=0}^{N-n-1} 
(N+1)^{n+1}t^\alpha t_{\alpha+n+1},\no \\
&&\hskip-8mm L_0=\sum_{\alpha=0}^N \sum_{m=0}^\infty 
(b_\alpha+m)t_m^\alpha\d_{\alpha,m}
+(N+1)\sum_{\alpha=0}^{N-1} \sum_{m=0}^\infty t_m^\alpha\d_{\alpha+1,m-1}
+\frac{1}{2\lambda^2}\sum_{\alpha=0}^{N-1} (N+1)t^\alpha t_{\alpha+1}\no\\
&&-\frac{1}{48}(N-1)(N+1)(N+3), 
\label{l0}\\
&&\hskip-8mm L_{-n}=\sum_{m=0}^\infty\sum_{\alpha=0}^N\sum_{j=0}^{N-\al}
(N+1)^jA_\alpha^{(j)}(m,n) t_{m+n+j}^\alpha\d_{\alpha+j,m} \hskip25mm n\ge 1
\label{l-n} \\
&&+{1 \over 2\lambda^2}\sum_{\alpha=0}^N \sum_{j=0}^{N-\al}
\sum_{m=0}^{n+j-1}(N+1)^jB_\alpha^{(j)}(m,n) t^\alpha_{n-m+j-1}t_{\alpha+j,m}.
\no
\ea
Here $\d_{\al,n}=\d/\d t_{n}^{\al}$ and $b_{\al}$ is related to the 
degree $q_{\al}$ of the cohomology class 
$\omega_{\al}$ as
\be
b_\alpha\equiv q_\alpha-\frac{1}{2}(N-1),\qquad
\omega_{\al}\in H^{q_{\al},\,q_{\al}}(CP^N), \qquad q_{\al}=0,1,\cdots, N.
\ee
The indices are raised and
lowered by the intersection form
\be
\eta^{\alpha\beta}=\int_M \omega_{\al}\wedge \omega_{\bt}.
\ee
$\eta^{\alpha\beta}=\delta_{\al+\bt,N}$ in $CP^N$. 
Note that $q^{\al}=N-q_{\al}, \hskip1mm b_{\al}+b^{\al}=1$.
$\lambda$ is the string coupling constant.

The coefficients functions are defined as\footnote{We have slightly 
changed the
normalization of the parameters; $t_n^\alpha$ is $n!$ times the one  
used in \cite{EHXa}.}
\ba
&&A_\alpha^{(j)}(m,n)\equiv(-1)^j\,\,\,
\frac{\Gamma(b_\alpha+m+j+1)}{\Gamma(b_\alpha+m+j+n)}
\sum_{1\le \ell_1\le\ell_2\le\cdots\le\ell_j\le n-1}
\hskip-1mm\bigg(\prod_{i=1}^j \frac{1}{b_\alpha+m+j+\ell_i}\bigg), 
\label{af}\\
&&B_\alpha^{(j)}(m,n) \equiv  (-1)^j
\frac{\Gamma(b^\alpha-j)\,\,\Gamma(b_\alpha+j)}
{\Gamma(b^\alpha+m-j)\,\Gamma(b_\alpha+n-m+j-1)}\no\\
&&\quad\times\sum_{0\le \ell_1\le\ell_2\le\cdots\le\ell_j\le n-2}
\bigg(\prod_{i=1}^j \frac{1}{b_\alpha+j-m+\ell_i}\bigg),\label{bf}
\\
&&C_\alpha^{(j)}(m,n)\equiv
\frac{\Gamma(b_\alpha+m+n+1)}{\Gamma(b_\alpha+m)}
\sum_{m\le \ell_1<\ell_2<\cdots<\ell_j\le m+n}
\bigg(\prod_{i=1}^j \frac{1}{b_\alpha+\ell_i}\bigg), 
\label{cf}\\
&& \no \\
&&D_\alpha^{(j)}(m,n)\equiv 
\frac{\Gamma(b^\alpha+m+1)\,\Gamma(b_\alpha+n-m)}
{\Gamma(b^\alpha)\,\,\Gamma(b_\alpha)}\no\\
&&\quad\times\sum_{-m-1\le \ell_1<\ell_2<\cdots<\ell_j\le n-m-1}
\bigg(\prod_{i=1}^j \frac{1}{b_\alpha+\ell_i}\bigg).
\label{df}\ea

By a somewhat lengthy calculation it is possible to show that
the above operators (\ref{ln}--\ref{l-n})
form the Virasoro algebra
\be
[L_n,L_m]=(n-m)L_{n+m}+{N+1 \over 12}(n^3-n)\delta_{n+m,0}, 
\qquad n,m \in {\bf Z}. 
\ee

The Virasoro condition is given by
\be
L_nZ=0, \hskip50mm n\ge -1
\ee
where $Z$ is the partition function of the topological $\sigma$-model
with the target space $CP^N$ and is related to the free energy as
$Z=\exp F=\exp(\sum_{g=0}\lambda^{2g-2}F_g)$. $F_g$ denotes the genus$=g$
free energy.
We have seen in \cite{EHXa} that the Virasoro
conditions at genus $g=0$ reproduce the associativity equations 
of quantum cohomology \cite{KM} and at genus=1 reproduce the 
elliptic Gromov-Witten invariants obtained in \cite{CH}, \cite{Getz}.

Let us now assemble the parameters $\{t_{\al,n}\}$ and construct 
a basic set of free scalar fields 
\ba
&&\phi^{\alpha}(z)=\sum_{n\ge0}\frac{\Gamma(b_\al)}{\Gamma(b_\al+n+1)}\,\,
t^\alpha_n\, z^{n+b_\al}-\sum_{n\ge0}\frac{\Gamma(b^\al+n)}
{\Gamma(b^\al)}\,\, z^{-n-b^\al}\,\,\partial_n^\al,\\
&&\hskip-3mm =-\sum_{n\in Z}\frac{\Gamma(b^{\al}+n)}
{\Gamma(b^{\al})}a_n^{\al}z^{-n-b^{\al}}, \hskip5mm 
a_n^{\al}\equiv \partial_n^{\al}, \hskip1mm
a_{-n-1}^{\al}\equiv(-1)^{n}t_n^{\al}, \,\, n\ge 0
\no \ea
Here $z$ is an arbitrary complex parameter.
It is easy to see that the 
scalar fields $\{\phi^{\al}\}$ have a canonical OPE. In terms of the currents,
$j^{\al}(z)=\partial_z\phi^{\al}(z)$ it is given by
\be
\la j_{\al}(z)j^\beta(w)\ra=\frac{\delta_{\al}^{\beta}}{(z-w)^2}
\bigg[ b_\al(\frac{z}{w})^{b^\al}+b^\al(\frac{w}{z})^{b_\al}\bigg]
\approx {\delta_{\al}^{\beta} \over (z-w)^2}+\mbox{constant}.
\label{OPEa}
\ee
The stress tensor is defined as the generating function of 
the Virasoro operators 
\be
T(z)=\sum_{n \in {\bf Z}} L_n z^{-n-2}.
\ee
We need a generalization of the currents $j^{\al}(z)$ 
by an additional parameter $\epsilon$
\be
\hskip-3mm
j^{\alpha}(z;\epsilon)=\sum_{n\in Z}\frac{\Gamma(b^{\al}+n+1-\epsilon)}
{\Gamma(b^{\al}-\epsilon)}\,\,a_n^{\al}z^{-n-b^{\al}-1},\hskip2mm
j_\alpha(z,-\e)=
\sum_{n\in Z}\frac{\Gamma(b_\al+n+1+\e)}
{\Gamma(b_\al+\e)}\,\, a_{\al,n}z^{-n-b_\al-1}.
\ee
We then find that the stress tensor can be represented as
\ba
&&\null\hskip-10mm T(z)=\sum_{\al,\bt=0}^N
\frac{1}{2}\exp\Big({\cal C}\partial_{\epsilon}
\Big)_{\hskip-1mm\al}^{\hskip1mm\bt}\hskip-1mm\left.
\left[\frac{\Gamma(b_{\bt}+\e)}
{\Gamma(b_{\al}+\e)}:j^{\al}(z;\e)j_{\bt}(z;-\e):\right]\right|_{\e=0} 
+\frac{1}{4}(\sum_\al b^\al b_\al)z^{-2}
\label{stressa} \\
&&\hskip-10mm=\sum_{\al,\bt=0}^N
\sum_{m,\ell}\frac{1}{2}\exp\Big({\cal C}\partial_{\epsilon}
\Big)_{\hskip-1mm\al}^{\hskip1mm\bt}\hskip-1mm
\left.\left[(-1)^{m+1}\frac{\Gamma(b_{\bt}+\ell+1+\e)}
{\Gamma(b_{\al}-m-1+\e)}\right]\right|_{\e=0}
\hskip-5mm:a_{m}^{\al}a_{\ell,\bt}:z^{-m-\ell-b^{\al}-b_{\bt}-2}
+\frac{1}{4}(\sum_\al b^\al b_\al)z^{-2}. \no
\ea
Here ${\cal C}$ denotes the matrix of the first-Chern class
\be
{\cal C}_{\al}^{\hskip2mm\bt}=\int_M c_1\wedge\omega_{\al}\wedge\omega^{\bt}.
\ee
${\cal C}_{\al}^{\hskip2mm\bt}=(N+1)\delta_{\al+1,\bt}$
in $CP^N$. Derivatives in $\epsilon$ of the ratio of gamma
functions in (\ref{stressa}) reproduce
the coefficient functions $A,B,C,D$ in
(\ref{af}--\ref{df}).

Since the stress tensor involves derivatives in $\e$, 
the basic currents $j^\alpha$ themselves (except $j^P$) do not have a good 
conformal property. Instead we may consider new currents defined by
\ba
&&J^{\al}(z)=\sum_{\bt=0}^{\al}
\exp\Big({\cal C}\partial_{\epsilon}\Big)_{\bt}^{\hskip2mm\al}
\Big(\frac{\Gamma(b_{\al}+\epsilon)}{\Gamma(b_{\bt}+\epsilon)}
j^{\bt}(z;\epsilon)\Big)|_{\epsilon=0} \no \\
&&\hskip3mm=\sum_{\bt=0}^{\al}\left.\Big(\exp{\cal C}\partial_{\epsilon}
\Big)_{\bt}^{\hskip2mm\al}\frac{(-1)^{m+1}\Gamma(b_{\al}+\e)}
{\Gamma(b_{\bt}-m-1+\e)}\right|_{\e=0}\hskip-3mm a_m^{\bt}z^{-m-b^{\bt}-1}.
\ea
After some algebra we find the following OPE 
\ba
&&\hskip-3mm\la J_{\al}(z)J^{\bt}(w)\ra=
\left({1 \over 1-{\cal C}}\right)_{\al}^{\hskip2mm\bt}
{1 \over (z-w)^3}
\left[z\Big(b_{\al}({w \over z})^{b_{\al}-1}+b^{\al}({w \over z})^{b_{\al}}
\Big)-w\Big(b_{\bt}({w \over z})^{b_{\bt}-1}+b^{\bt}({w \over z})^{b_{\bt}}
\Big)\right] \no \\
&&\hskip10mm \approx \left({1 \over 1-{\cal C}}\right)_{\al}^{\hskip2mm\bt}  
\left({1 \over (z-w)^2}+
{b_{\al}b^{\al}-b_{\bt}b^{\bt} \over 2(z-w)}{1 \over w}\right) \hskip20mm
\bt \ge \al. 
\label{OPEb} 
\ea
$\la J_{\al}(z)J^{\bt}(w)\ra$ is regular for $\al>\bt$.
These currents $J_{\al}$ have a good transformation law with respect to the 
stress tensor 
\ba
T(z)J^{\al}(w)&=&{1 \over (z-w)^2}({w \over z})^{b_\al}J^{\al}(z) 
+\frac{1}{w(z-w)}\sum_{\bt}J^{\beta}(z)\, 
\left({1 \over {1-\cal C}}\right)_{\hskip-1mm\bt}^{\hskip1mm\al}\,
b_\beta (\frac{z}{w})^{b^\beta}\label{tope}\\
 &\approx & {1 \over (z-w)^2}J^{\al}(w)
+{1 \over (z-w)}\left(\partial_w J^{\al}(w)+\sum_{\bt}\left({{\cal C} \over
1-{\cal C}}\right)_{\hskip-1mm\bt}^{\hskip1mm\al}
b_{\bt}{J^{\bt}(w) \over w}\right).
\no\ea
In fact it is possible to show that the stress tensor itself is expressed 
as a bilinear form in terms of
$J_{\al}$
\be
T(z)={1 \over 2}\sum_{\al,\bt}:G^{\al\bt}J_{\al}(z)J_{\bt}(z):
+{1 \over 4}(\sum_{\al}b_{\al}b^{\al})z^{-2}, \hskip4mm 
G^{\al\bt}=\eta^{\al\bt}-{\cal C}^{\al\bt}.
\label{tbilinear}\ee
Derivation of formulas (\ref{OPEb}),(\ref{tope}),(\ref{tbilinear}) is a bit
involved and will not be not presented here. One may, however, easily check 
the consistency among these equations.

We may further redefine the fields and bring the OPE (\ref{OPEb}) into the
canonical form (\ref{OPEa}). For instance, in the case of $CP^2$ we may
introduce scalar fields $\varphi^{\al}(z) \, (\al=P,Q,R)$ by
\ba
&&\hskip-3mm \partial_z\varphi^P(z)=J^P(z), \hskip1mm
\partial_z\varphi^Q(z)-{3 \over 2z}\varphi^P(z)
=J^Q(z)-{3 \over 2}J^P(z), \\ 
&&\partial_z\varphi^R(z)=J^R(z)-{3 \over 2}J^Q(z)-{9 \over 8}J^P(z).
\no \ea
Then we have
\be
\partial_z\varphi_{\al}(z)\partial_w\varphi^{\bt}(w) 
\approx {\delta_{\al}^{\bt} \over (z-w)^2}+\mbox{constant}
\ee
and
\be
T(z)={1 \over 2}\Big(\partial_z\varphi^Q(z)-{3 \over 2z}\varphi^P(z)\Big)^2
+\partial_z\varphi^P(z)\partial_z\varphi^R(z)-{5 \over 16}{1 \over z^{2}}.
\ee
The above realization is perhaps the closest to the standard conformal 
field theories. Stress tensor contains anomalous non-derivative 
terms proportional to $\varphi^P(z)$.
The conformal algebra is still maintained in the presence
of such terms. Similar construction may be given for general $CP^N$. \\

\bigskip

\noindent{\large \bf 3. Holomorphic Virasoro Algebra and Cubic Three-fold} \\

\bigskip

In the case of projective spaces cohomology classes only of the analytic type
$H^{q,q}$ appear. These are even-dimensional classes and hence the 
corresponding fields
are necessarily bosonic. In \cite{EHXa} we have considered 
manifolds with only analytic cohomology classes. In these cases 
the general form of the $L_0$ operator is given by \cite{Ho}
\ba
L_0&=&
\sum_{m=0}^{\infty}(m+b_{\al})t_m^{\alpha}{\partial\over
\partial t_m^{\alpha}}
+\sum_{m=1}^{\infty} ({\cal C})_{\alpha}^{\hskip2mm \bt}
t_m^{\alpha}{\partial
\over \partial t_{m-1}^{\beta}}
+{1\over 2\lambda^2}({\cal C})_{\alpha\beta}t^{\alpha}t^{\beta}\nonumber\\
&&
+{1\over 24}\left({3-n\over2}\int_M c_n-\int_M c_1\wedge c_{n-1}
\right)
\ea
Here $n$ is the dimension of $M$ and $c_j$ denotes its $j$-th Chern class.
The constant term of $L_0$ (the "vacuum energy") has a specific 
dependence on the manifold. In our construction the
Virasoro algebra holds if the relation 
\ba
{1 \over 4}\sum_{\al}b^{\al}b_{\al}={1 \over 4}\sum_{q}
\Big(n-q-{(n-1) \over 2}\Big)\Big(q-{(n-1) \over 2}\Big)h^{q,q}
={1 \over 24}\Big({3-n \over 2}\int_M c_n
-\int_M c_1\wedge c_{n-1}\Big)
\label{curious}
\ea
is satisfied ((\ref{curious}) follows from the commutator $[L_1,L_{-1}]=2L_0$).
In \cite{Lib,Bori} it was shown that in fact (\ref{curious}) holds if 
the manifold has only analytic cohomology classes. 

In the general case when
"off-diagonal" cohomology classes $H^{p,q},\,p\not=q$ are present
one has \cite{Bori}
\be
{1 \over 4}\sum_{p,q}(n-p-{n-1 \over 2})(p-{n-1 \over 2})h^{p,q}(-1)^{p+q}
={1 \over 24}\Big({3-n \over 2}\int_M c_n-\int_M c_1\wedge c_{n-1}\Big).
\label{Brisovb}
\ee
Comparison of (\ref{Brisovb}) with (\ref{curious}) suggests that one should
define the factor $b_{\al}$ in the Virasoro operators as 
$b_{\al}\equiv p-(n-1)/2$ for a cohomology class $\omega\in H^{p,q}(M)$.
One may as well define $b_{\al}\equiv q-(n-1)/2$ since $h^{p,q}$ is symmetric 
in $p,q$ and the LHS of the above equation remains invariant under $p,q$ 
interchange. Let us call these values of $b_{\al}$ as the holomorphic and
anti-holomorphic dimensions.

Based on the above observation S.Katz \cite{Kz} 
has proposed to define holomorphic
(anti-holomorphic) Virasoro operators where the factors $b_{\al}$ are given
by the holomorphic (anti-holomorphic) dimensions. The sign 
factor $(-1)^{p+q}$ in (\ref{Brisovb}) takes care of the
statistics in the case of cohomology classes with $p+q$=odd 
which are described by 
anti-commuting variables.

Let us now consider specific manifolds with off-diagonal cohomology classes.
In order to fix our discussions we consider the case of hypersurfaces in
complex projective spaces. In this case there appear extra classes,
primitive cohomologies $\{\omega_i\in H^{p_i,q_i}\}$ 
at the middle dimension
$p_i+q_i=n$ (Lefschetz hyperplane theorem).
These classes are decoupled 
from the K\"{a}hler subring
generated by $\{\omega_{\al}\in H^{q_{\al},q_{\al}} 
\,(q_{\al}=0,1,\cdots,n)\}$.
 Also the matrix of first-Chern class
vanishes for primitive cohomologies, ${\cal C}_{i\al}={\cal C}_{ij}=0$.
Thus they contribute additional diagonal terms to the stress tensor.

For the sake of definiteness let us consider the case of
odd-dimensional hypersurface. Then all the primitive classes 
are odd dimensional 
and their couplings $\{u_{i,n}\}$ become anti-commuting variables. 
Let us introduce $BC$ ghosts for each pair of primitive classes
$\omega_i,\,\omega^i$,
\ba
&&B_{i}(z)=\sum_{n\ge 0}{\Gamma(b^i) \over \Gamma(b^i+n+1)}u_{i,n}
z^{n+b^i}+\sum_{n\ge 0}{\Gamma(b_i+n) \over \Gamma(b_i)}
\partial_{i,n}z^{-n-b_i}, \\
&&C^{i}(z)=\sum_{n\ge 0}{\Gamma(b_i) \over \Gamma(b_i+n)}u_{n}^i
z^{n+b_i-1}+\sum_{n\ge 0}{\Gamma(b^i+n+1) \over \Gamma(b^i)}
\partial_{n}^iz^{-n-b^i-1}.
\label{bc}
\ea
Here $b_i$ denotes the holomorphic dimension $p_i-(\mbox{dim M}-1)/2$ 
of $\omega_i$ and
$\partial_{i,n}=\partial/\partial u_{i,n}$. OPE's are given by
\be
\la B_{i}(z)C^{j}(w)\ra=\delta_i^j{1 \over z-w}({z \over w})^{b^i}
\approx \delta_{i}^{j}\left({1 \over (z-w)}+{b^i \over w}\right).
\ee
Stress tensor of the BC system is defined as  
\be
T_{ghost}(z)=\sum_i :\partial_zB_i(z)C^i(z):
-\frac{1}{2}(\sum_{i}b_i\, b^i)z^{-2}.
\ee
As is well-known, a pair of BC ghosts (of spin 0 and 1)
contributes $-2$ to the central charge 
and hence the odd-dimensional cohomologies contribute --(number of odd classes)
in total. Thus the Virasoro central charge equals
\be
c=\mbox{number of even classes}-\mbox{number of odd classes}=\chi(M)
\ee
for a general manifold $M$ where $\chi(M)$ is its Euler characteristic.

In the case of degree $k$ hypersurface in $CP^N$ (we denote it as
$M_N^k$) the holomorphic
Virasoro operators are given by 
\ba
&&L_{n}=\sum_{m=0}\sum_{\alpha,j}
(N+1-k)^{j}C_{\alpha}^{(j)}(m,n)
t_{m}^{\alpha}\partial_{\alpha+j,m+n-j} \no \\
&&+\frac{\lambda^{2}}{2}\sum_{m=0}^{n-j-1}\sum_{\alpha,j}
(N+1-k)^{j}D_{\alpha}^{(j)}(m,n)\partial_{m}^{\alpha}
\partial_{\alpha+j,n-m-j-1}
\nonumber\\
&&+\sum_{m=0}\sum_{i}C_{i}^{(0)}(m,n)
u_{m}^{i}\partial_{i,m+n}+\sum_{m=0}\sum_{\bar{i}}C_{\bar{i}}^{(0)}(m,n)
u_{m}^{\bar{i}}\partial_{\bar{i},m+n} \no \\
&&+\lambda^{2}\sum_{m=0}^{n-1}\sum_{i}D_{i}^{(0)}(m,n)
\partial_{m}^{i}\partial_{i,n-m-1} 
\label{lnb} \\
&&+\frac{1}{2\lambda^{2}}(N+1-k)^{n+1}\sum_{\alpha}
t^{\alpha}t_{\alpha+n+1}, \hskip40mm n\ge 1, \no \\
&& \no \\
&&L_{0}=\sum_{m=0}\sum_{\alpha}
(b_{\alpha}+m)t_{m}^{\alpha}\partial_{\alpha,m}
+(N+1-k)\sum_{m=1}\sum_{\alpha}t_{m}^{\alpha}\partial_{\alpha+1,m-1} \no \\
&&+\frac{1}{\lambda^{2}}(N+1-k)\sum_{\alpha}
t^{\alpha}t_{\alpha+1}
+\sum_{m=0}\sum_{i}
(b_{i}+m)u_{m}^{i}\partial_{i,m}
+\sum_{m=0}\sum_{\bar{i}}
(b_{\bar{i}}+m)u_{m}^{\bar{i}}\partial_{\bar{i},m}\no  \\
&&+\frac{1}{24}(\frac{4-N}{2}\int_M c_{N-1}-\int_M c_{1}\wedge c_{N-2}),
\label{l0b}\\
&& \no \\
&&L_{-n}=\sum_{m=0}\sum_{\alpha,j}
(N+1-k)^jA_\alpha^{(j)}(m,n) t_{m+n+j}^\alpha\d_{\alpha+j,m} \no \\
&&+{1 \over 2\lambda^2}\sum_{\alpha,j}
\sum_{m=0}^{n+j-1}(N+1-k)^j 
B_\alpha^{(j)}(m,n) t^\alpha_{n-m+j-1}t_{\alpha+j,m}
\label{l-nb} \\
&&+\sum_{m=0}\sum_{i}A_i^{(0)}(m,n) u_{m+n}^i\d_{i,m}
+\sum_{m=0}\sum_{\bar{i}}A_{\bar{i}}^{(0)}(m,n) u_{m+n}^{\bar{i}}
\d_{\bar{i},m} \no \\
&&+{1 \over \lambda^2}\sum_{i}\sum_{m=0}^{n+j-1}
B_i^{(0)}(m,n) u_{i,m}u^i_{n-m-1} \hskip40mm n\ge 1 \no.
\end{eqnarray}   
Note that $(N+1-k)$ is the first Chern class of $M_N^k$.

It is known that two-dimensional hypersurfaces are exceptional and have 
only analytic
cohomology classes. It is also known that the degree of hypersurfaces 
must be greater than 2 to possess non-analytic classes.
Thus the simplest 
example with off-diagonal classes appears to be $M_4^3$, 
cubic hypersurface in $CP^4$.
We have tested the predictions of the holomorphic
Virasoro operators (\ref{l0b}-\ref{l-nb}) upto genus$=1$ for $M_4^3$.
It turns out that they agree precisely with those obtained by the 
associativity relations and genus=1 recursion relation \cite{Getz}.

Non-vanishing Hodge numbers of $M_4^3$ are given by 
$h^{q,q}(M_4^3)=1 \,(q=0,1,2,3)$ and $h^{2,1}(M_4^3)=h^{1,2}(M_4^3)=5$.
We denote the primary fields 
as P,Q,R,S or $\Op_{\al}\,(\al=0,1,2,3)$ (K\"{a}hler
sector) and $\Op_i\,(i=1,\cdots,5),\, \Op_{\bar{i}}\,(\bar{i}=1,\cdots,5)$ 
(primitive sector). The intersection parings are given by
\begin{equation}
\langle PPS \rangle =3 ,\; \langle PQR \rangle =3,\;
\langle P{\cal O}_{i}{\cal O}_{\bar{j}}\rangle
=-\langle P{\cal O}_{\bar{j}}{\cal O}_{i}\rangle \equiv -d_{i\bar{j}}=
-\delta_{i\bar{j}}
\label{metric3}
\end{equation}
Minus sign on the third term comes from the anti-commuting nature of 
Grassmann variables. We have chosen a basis so that the metric
of the primitive sector $d_{i\bar{j}}$
is diagonalized.

At genus=0 the Virasoro condition reduces to the associativity equation and
automatically reproduces the correct Gromov-Witten invariants.
Thus we examine the Virasoro conditions at genus=1. 
Genus 0 and 1 free energies of the $M_4^3$ model have the following
instanton expansions
\begin{eqnarray}
&&F_{0}= \frac{3}{2}(t^{P})^{2}t^{S}+3t^{P}t^{Q}t^{R}
+ \frac{3}{6}(t^{Q})^{3}+t^{P}d_{i\bar{j}}u^{i}u^{\bar{j}}+
f(t^{Q},t^{R},t^{S},u^{i},u^{\bar{i}}); \\
&&f(t^{Q},t^{R},t^{S},u^{i},u^{\bar{i}})=
\sum_{{\scriptstyle d,a,b},{\{m_{i}\},\{m_{\bar{i}}\}}}
 N_{d,a,b,\{m_{i}\},\{m_{\bar{i}}\}}^{0}\frac{(t^{R})^{a}(t^{S})^{b}}{a!b!}
\prod_{i}(u^{i})^{m_{i}}(u^{\bar{i}})^{m_{\bar{i}}}\exp{dt^{Q}},\nonumber\\
&&\hskip40mm \mbox{where} 
\hskip5mm 
a+2b+\sum_{i=1}^{5}m_{i}=2d, \hskip2mm  a+2b+\sum_{\bar{i}=1}^{5}m_{\bar{i}}=2d
\label{inst0}
\ea
and
\ba
&&F_{1}=-\frac{1}{2}t^{Q}+
g( t^{Q},t^{R},t^{S},u^{i},u^{\bar{i}} );
\\
&&g(t^{Q},t^{R},t^{S},u^{i},u^{\bar{i}})=
\sum_{{\scriptstyle d,a,b},{\{m_{i}\},\{m_{\bar{i}}\}}}
 N_{d,a,b,\{m_{i}\},\{m_{\bar{i}}\}}^{1}\frac{(t^{R})^{a}(t^{S})^{b}}{a!b!}
\prod_{i}(u^{i})^{m_{i}}(u^{\bar{i}})^{m_{\bar{i}}}\exp{dt^{Q}},\no \\
&&\hskip40mm \mbox{where} \hskip5mm 
a+2b+\sum_{i=1}^{5}m_{i}=2d, \hskip2mm a+2b+\sum_{\bar{i}=1}^{5}m_{\bar{i}}=2d.
\label{inst1}
\ea

Let us first look at
the $L_1$ equation,
\begin{eqnarray}
&&L_1Z=\left[\sum_{\alpha=0}^{3}\sum_{n=0}^{\infty}
(n+b_{\alpha})(n+b_{\alpha}+1)
t_{n}^{\alpha}\partial_{\alpha,n+1}
+2\sum_{\alpha=0}^{3}\sum_{n=0}^{\infty}(2n+2b_{\alpha}+1)t_{n}^{\alpha}
\partial_{\alpha+1,n}\right.\nonumber\\
&&\left.+4\sum_{\alpha=0}^{3}\sum_{n=1}^{\infty}t_{n}^{\alpha}
\partial_{\alpha+2,n-1}
+{\lambda^2 \over 2}\sum_{\alpha=0}^{3}
b_{\alpha}b^{\alpha}\partial_{\alpha}\partial^{\alpha}
+{2 \over \lambda^2}\sum_{\alpha=0}^{3}t^{\alpha}t_{\alpha+2}
+\lambda^2\sum_{i=1}^5 b^ib_i\partial^i\partial_i\right. \\
&&\left.+\sum_{i=1}^{5}\sum_{n=0}^{\infty}(n+b_{i})(n+b_{i}+1)
u_{n}^{i}\partial_{i,n+1}+\sum_{\bar{i}=1}^{5}
\sum_{n=0}^{\infty}(n+b_{\bar{i}})(n+b_{\bar{i}}+1)
u_{n}^{\bar{i}}\partial_{\bar{i},n+1}\right]Z=0. \no
\end{eqnarray}
In the small phase space ($\{t^{\al}_n\}=\{u^i_{n}\}=\{u^{\bar{i}}_{n}\}=0$ 
for $n\ge 1$ except $t^P_1=-1$) and at the genus=1 level
the above $L_1$ condition reads as
\begin{eqnarray}
&&t^{P}(2t^{R}g_{R}+4t^{S}g_{S}+\sum_i 2u^{i}g_{i}-2\cdot2\cdot g_{Q}
+2\cdot{1 \over 2}-1)=0,
\label{tpterm} \\
&&4g_{R}-4t^{R}g_{S}
=\frac{1}{18}t^{R}f_{RQR}+\frac{1}{6}
t^{S}f_{SQR}+\frac{1}{18}u^{i}f_{iQR}
-\frac{1}{18}f_{QQR}
+\frac{1}{6}t^{R}f_{Ri\bar{j}}d^{i\bar{j}}+
\frac{1}{2}t^{S}f_{Si\bar{j}}d^{i\bar{j}}  \no \\
&&+\frac{1}{6}u^{k}f_{ki\bar{j}}d^{i\bar{j}}
-\frac{1}{6}f_{Qi\bar{j}}d^{i\bar{j}} -\frac{1}{6}
(2t^{R}f_{RR}+6t^{S}f_{SR}
+2u^{i}f_{iR}-{2}f_{QR})+2u^iu_ig_S  \label{l1rec} \\
&&+{1 \over 3}(2t^{R}f_{RQ}+6t^{S}f_{SQ}+2u^{i}f_{iQ}-{2}f_{QQ})g_{R}
+{1 \over 3}(2t^{R}f_{RR}+6t^{S}f_{SR}+2u^{i}f_{iR}-{2}f_{QR})g_{Q} 
\no \\
&&+(2t^{R}f_{Ri}+6t^{S}f_{Si}+2u^{k}f_{ki}-{2}f_{Qi})g_{\bar{j}}d^{i\bar{j}}
-(2t^{R}f_{R\bar{j}}+6t^{S}f_{S\bar{j}}+2u^{k}f_{k\bar{j}}
-{2}f_{Q\bar{j}})g_id^{i\bar{j}}.\nonumber
\end{eqnarray}
Here $g_Q\equiv\partial g/\partial t^Q,f_{RQ}=\partial^2 f
/\partial t^R\partial t^Q$ etc. and the topological recursion relations at 
genus=0 and 1 have been used.
Note that eq.(\ref{tpterm}) is satisfied because of the condition
on the instanton expansion $a+2b+\sum{m_i}=2d$ of the function $g$ 
(\ref{inst1}). 

We determine instanton numbers $N^1_{d,a,b,\{m_i=0\},\{m_{\bar{i}}=0\}}
\equiv N^1_{d,a,b}$.
Anti-commuting parameters $\{u_i\}$ are eventually set to zero in 
deriving these instanton numbers. Then terms of the form
$u^if_{Ri}$ drop out, however, there are "loop" corrections of the 
type $f_{Ri\bar{j}}d^{i\bar{j}}$ which give non-vanishing contributions.
In order to evaluate these terms we consider two-point functions
\begin{eqnarray}
&&\langle{\cal O}_{i}{\cal O}_{\bar{i}}\rangle\;|_{\{u^j=0\}}=
-t^{P}+h(t^{Q},t^{R},t^{S})\nonumber\\
&&h(t^{Q},t^{R},t^{S})=\sum_{d=1}^{\infty}\sum_{a+2b=2d}\tilde{N}_{d,a-1,b}^{0}
\frac{(t^{R})^{a-1}(t^{S})^{b}}{(a-1)!b!}\exp(dt^{Q}).
\end{eqnarray}
(It is possible to show inductively in degree $d$ that 
$\langle{\cal O}_{i}{\cal O}_{\bar{j}}\rangle\;|_{\{u^k=0\}}=0$ 
for $i \not=j$).
Expansion coefficients $\tilde{N}_{d,a-1,b}^{0}$ are determined from the 
associativity relation at genus=0 
\be
h_{R}=\frac{1}{3}f_{SQQ}-\frac{1}{3}h_{Q}f_{RQQ}-\frac{1}{3}h_{R}f_{QQQ}
-(h_{Q})^{2}.
\label{aux}
\end{equation}
(\ref{aux}) leads to a relation
\begin{eqnarray}
&&\hskip-3mm\tilde{N}_{d,a-1,b}^{0}=\frac{1}{3}d^{2}N_{d,a-2,b+1}^{0} \no \\
&&\hskip-3mm-\frac{1}{3}\sum_f\sum_j\sum_{\hskip1mmi+2j=2f}
\tilde{N}_{f,i-1,j}^{0}N_{d-f,a-i,b-j}^{0}
\left(f(d-f)^{2}{a-2\choose i-1}{b\choose j}
+(d-f)^{3}{a-2\choose i-2}{b\choose j}\right)\nonumber\\
&&\hskip-3mm-\sum_f\sum_j\sum_{\hskip1mmi+2j=2f}
\tilde{N}_{f,i-1,j}^{0}\tilde{N}_{d-f,a-i-1,b-j}^{0}
f(d-f){a-2\choose i-1}{b\choose j}.
\end{eqnarray}
(here $N^0_{d,a,b,\{m_i=0\},\{m_{\bar{i}}=0\}}\equiv N^0_{d,a,b}$). From
(\ref{l1rec}) we obtain the recursion formula for the number of elliptic curves
\begin{eqnarray}
&&N_{d,a,b}^{1}-(a-1)N_{d,a-2,b+1}^{1}\nonumber\\
&&=\frac{1}{72}\Big((d-6)N_{d,a,b}^{0}+15\tilde{N}_{d,a-1,b}^{0}\Big)
(d+b-1) \\
&&+\frac{1}{6}\sum_{\stackrel{\scriptstyle  u+c=a}{v+e=b}}
\sum_{f+g=d}N_{f,u,v}^{0}N_{g,c,e}^{1}
\Big({a-1\choose u}{b\choose v}(f+v) f+
{a-1\choose u-1}{b\choose v}(f+v-1) g\Big). \nonumber
\end{eqnarray}
The above relation can not completely determine $N_{d,a,b}^{1}$.
Its boundary values ($N^1_{d,a,b}$ at $a=0$ or $b=0$) 
are left undetermined. In order to fix them we 
have to invoke the $L_2$ condition. In the small phase space
$L_2$ constraint leads to
\begin{eqnarray}
&&8g_{S}=\frac{1}{18}(2-d_{i\bar{j}}d^{i\bar{j}})
(3t^{R}f_{RS}+12t^{S}f_{SS}-2f_{QS})-{4 \over 3}f_{QS}\no \\
&&-{1 \over 3}(8t^{R}f_{RS}+6t^{S}f_{SS}-2f_{QS}-4f_{RR}-2f_{S})\nonumber\\
&&+\frac{1}{2}t^{R}f_{SQR}+\frac{3}{2}t^{R}f_{Si\bar{j}}d^{i\bar{j}}
-{1 \over 3}f_{RRQ}-f_{Ri\bar{j}}d^{i\bar{j}}
+{1 \over 3}f_{RR}-2f_{S}\nonumber\\
&&+\frac{1}{54}(3t^{R}f_{RQ}+12t^{S}f_{SQ}-2f_{QQ})f_{RQR}
+\frac{1}{54}(3t^{R}f_{RR}+12t^{S}f_{SR}-2f_{QR})f_{QQR} \no \\
&&+\frac{1}{18}(3t^{R}f_{RQ}+12t^{S}f_{SQ}-2f_{QQ})f_{Ri\bar{j}}d^{i\bar{j}}
+\frac{1}{18}
(3t^{R}f_{RR}+12t^{S}f_{SR}-2f_{QR})f_{Qi\bar{j}}d^{i\bar{j}}
\no \\
&&-{1 \over 2}\left({2 \over 9}(2t^{R}f_{RQ}+6t^{S}f_{SQ}-2f_{QQ})f_{RR}
+{1 \over 9}(2t^{R}f_{RR}+6t^{S}f_{SR}-2f_{QR})f_{QR}\right) \\
&&+{2 \over 3}g_{S}(3t^{R}f_{R}+12t^{S}f_{S}-4f_{Q})\nonumber\\
&&+g_{R}
\left({8 \over 3}t^{R}f_{SQ}-{4 \over 3}f_{QR}+{2 \over 3}f_{R}
+{2 \over 3}(3t^{R}f_{RR}+8t^{S}f_{SR}-4f_{QR})\right.\nonumber\\
&&\left.-{1 \over 3}t^{R}(3t^{R}f_{RS}+8t^{S}f_{SS}-4f_{QS})
+{1 \over 9}f_{RQ}(3t^{R}f_{RQ}+8t^{S}f_{SQ}-4f_{QQ})\right)
\nonumber\\
&&+g_{Q}\left(4t^{R}f_{RS}-{8 \over 3}f_{RR}
-{4 \over 3}f_{S}+{2 \over 3}(2t^{R}f_{RS}+6t^{S}f_{SS}-2f_{QS})\right.
\no \\
&&\left.+{2 \over 9}f_{RR}(2t^{R}f_{RQ}+6t^{S}f_{SQ}-2f_{QQ})
+{1 \over 9}f_{RQ}(2t^{R}f_{RR}+6t^{S}f_{SR}-2f_{QR})\right).
\no
\end{eqnarray}
Here the recursive formula eq.(2.12) of \cite{EHXb} has 
been used. 
The above $L_2$ equation can be converted into an extra relation on
the number of elliptic curves 
\begin{eqnarray}
&&8N_{d,a,b}^{1}=(\frac{1}{2}ad-\frac{1}{3}d-\frac{19}{6}a-4b+\frac{8}{3})
N_{d,a,b}^{0}+{1 \over 3}(5-d)N_{d,a+2,b-1}^{0}\nonumber\\
&&+\frac{15}{2}a\tilde{N}_{d,a-1,b}^{0}-5\tilde{N}_{d,a+1,b-1}^{0}\nonumber\\
&&+\sum_{f+g=d}\sum_{\stackrel{\scriptstyle i+k=a}{j+l=b}}
{a\choose i}{b-1\choose j}
\Big[(3i+12j-2f)\left(\frac{1}{54}(
fgN_{f,i,j}^{0}N_{g,k+2,l-1}^{0}+g^{2}N_{f,i+1,j}^{0}N_{g,k+1,l-1}^{0})
\right.\nonumber\\
&&\left.+\frac{5}{18}(fN_{f,i,j}^{0}\tilde{N}_{g,k+1,l-1}^{0}+
gN_{f,i+1,j}^{0}\tilde{N}_{g,k,l-1}^{0})\right)\nonumber\\
&&+(2i+6j-2f)(-\frac{1}{9}fN_{f,i,j}^{0}N_{g,k+2,l-1}^{0}
-\frac{1}{18}gN_{f,i+1,j}^{0}N_{g,k+1,l-1}^{0})\nonumber\\
&&+(3i+12j-4f)(\frac{2}{3}N_{f,i,j}^{0}N_{g,k,l}^{1})
\label{l2rec} \\
&&+\frac{8}{3}(i\cdot f)N_{f,i-1,j+1}^{0}N_{g,k+1,l-1}^{1}
-{4 \over 3}fN_{f,i+1,j}^{0}N_{g,k+1,l-1}^{1}
+{2 \over 3}N_{f,i+1,j}^{0}N_{g,k+1,l-1}^{1}\nonumber\\
&&+(3i+8j-4f)\cdot({2 \over 3}N_{f,i+1,j}^{0}N_{g,k+1,l-1}^{1}
-\frac{1}{3}kN_{f,i,j+1}^{0}N_{g,k,l-1}^{1})\nonumber\\
&&+4(i\cdot g)N_{f,i,j+1}^{0}N_{g,k,l-1}^{1}
-{8 \over 3}gN_{f,i+2,j}^{0}N_{g,k,l-1}^{1}
-\frac{4}{3}gN_{f,i,j+1}^{0}N_{g,k,l-1}^{1}\nonumber\\
&&+\frac{2}{3}g(2i+6j-2f)N_{f,i,j+1}^{0}N_{g,k,l-1}^{1}\Big]\nonumber\\
&&+\sum_{f+g+h=d}\sum_{\stackrel{\scriptstyle i+k+m=a}{j+l+n=b}}
{a\choose {i,k}}
{b-1\choose {j,l}}\left((3i+8j-4f)\frac{f\cdot g}{9}
N_{f,i,j}^{0}N_{g,k+1,l}^{0}N_{h,m+1,n-1}^{1}\right.\nonumber\\
&&\left.+(2i+6j-2f)\frac{2f\cdot h}{9}
N_{f,i,j}^{0}N_{g,k+2,l}^{0}N_{h,m,n-1}^{1}+(2i+6j-2f)\frac{g\cdot h}{9}
N_{f,i+1,j}^{0}N_{g,k+1,l}^{0}N_{h,m,n-1}^{1}\right). \no
\end{eqnarray}
(\ref{l1rec}) and (\ref{l2rec}) determine the elliptic invariants of
the $M_4^3$ model. The results are presented in Table 1. 
(In the above calculation
we have used the convention of the metric $\la PPS\ra=\la PQR\ra=3$ which
is natural from the point of view of ring structure of cohomology
classes. On the other hand from the point of view of enumerative geometry the
metric $\la PPS\ra=\la PQR\ra=1$ is preferred. These conventions are related to
each other simply by rescaling by a factor 3 of $t^R$ and $t^S$.
In Table 1 we present the numbers $N_{d,a,b}$ 
which are obtained after rescaling by a factor $3^{a+b}$.)
It is possible to
show that our predictions agree completely with
those obtained by using the $g=1$ recursion relation of \cite{Getz}.

We note that if we had used anti-holomorphic dimensions for $b_i$ in the
above computation, we
would have obtained equations where $i$ and $\bar{i}$ are everywhere
interchanged.
Results on $N^0_{d,a,b},N^1_{d,a,b}$ are of course left unchanged. \\

\bigskip

\noindent{\large \bf 4. Discussions} \\

\bigskip

In our previous article \cite{EHXa} we were mainly concerned with the
quantum cohomology of Fano varieties. It now appears, however, that
our construction is quite general and with a suitable modification
Virasoro algebra controls the
quantum cohomology of all smooth compact K\"{a}hler manifolds. 
It is an important
issue to find exactly the class of manifolds where Virasoro condition
works. 

We have seen that in the case of Fano varieties the structure of free fields
and Virasoro
operators are deformed significantly due to non-vanishing first Chern class.
In the case of manifolds with vanishing first Chern class,
however, mixing between different cohomology classes disappear and
the Virasoro operators become simply a sum of diagonal terms 
coming from each cohomology class. The 
free field representation in such a case
may provide a natural interpretation of
the appearance of an affine Lie algebra in the moduli 
space of Yang-Mills instantons \cite{Nak,VW}. 
In the case of the $K_3$ surface, for instance,
we may introduce 24 free bosonic fields corresponding to 24 (even) homology 
classes.
These fields will generate an affine Lie 
algebra $E_8\times E_8\times U(1)^8$ (at level 1)  
whose partition function is given by $\eta(\tau)^{-24}$. This coincides 
with the
formula by Vafa and Witten \cite{VW} 
for the partition function of $N=4$ 
supersymmetric Yang-Mills theory in $K_3$ surface.
It is interesting to see if our world sheet 
approach may also help explaining some of the results of string duality.\\

\bigskip

\noindent{\large \bf Acknowledgement}\\

\bigskip

We would like to thank K. Hori for his collaboration at the early stage of
this work and S. Katz for his communication.
We are grateful to E. Getzler for stimulating discussions and
E. Witten for his encouragement.

\begin{table}[p]
\caption{Instanton numbers in $M_{4}^{3}$ model}
\begin{center}
\begin{tabular}{|l|l|l|l|l|}
\hline
$d$&
 $a,b$&
 $N_{d,a,b}^{0}$&$N_{d,a,b}^{1}$&
$\frac{2d-2}{24}N_{d,a,b}^{0}+N_{d,a,b}^{1}$\\
\hline
 1&(0,1) & 6 & 0 & 0  \\
\hline
 1&(2,0)& 5 & 0 &0 \\
\hline
 2&(0,2)& 6 &$-(1/2)$  &0\\
\hline
 2&(2,1) & 14 &$-(7/6)$&0\\
\hline
 2&(4,0)& 27 &$ -(9/4)$  &0 \\
\hline 
 3&(0,3)& 24 &$-3$ &1 \\
\hline
 3&(2,2) & 92 & $-(43/3)$  & 1 \\
\hline
 3&(4,1) & 270 & $-43$  & 2   \\
\hline
 3&(6,0) & 855 & $-(275/2)$  & 5 \\
\hline
 4&(0,4) & 192  & $-21$  & 27   \\
\hline
 4&(2,3) & 1024 & $-199 $ & 57  \\
\hline
 4&(4,2) & 4104 & $-846 $ & 180  \\
\hline
 4&(6,1) & 16992 & $-3588$  & 660   \\
\hline
 4&(8,0) & 74982 & $-(32211/2)$ & 2640   \\
\hline
 5&(0,5) & 2400  & 4 & 804 \\
\hline
 5&(2,4) & 16400  & $-(8252/3)$  & 2716   \\
\hline
 5&(4,3) & 84024 & $-16371$  & 11637  \\
\hline
 5&(6,2) & 432612  & $-89013$  & 55191  \\
\hline
 5&(8,1) & 2324166  & $-493678$  & 281044  \\
\hline
 5&(10,0) & 13153185 & $-2871189$  &  1513206  \\
\hline                     
\end{tabular}
\end{center}
\end{table}

\end{document}